\documentclass[twocolumn,eqsecnum,aps,prd,showpacs,preprintnumbers,amsmath,amssymb,floatfix]{revtex4}
\usepackage{bm}
\usepackage{graphicx,epsf}
\usepackage{pstricks}
\usepackage{amssymb}

\newcommand\ee{\end{equation}}

\newcommand\be{\begin{equation}}

\newcommand\eea{\end{eqnarray}}

\newcommand\bea{\begin{eqnarray}}

\newcommand\sub[1]{_{\rm #1}}

\newcommand\eq[1]{Eq.~(\ref{#1})}

\newcommand\fig[1]{Fig.~\ref{#1}}

\begin{document}
\title{The Tensor Desert}
\author{Laila Alabidi\footnote{electronic address: l.alabidi@lancaster.ac.uk}}
\affiliation{Physics Department, Lancaster University,LA1 4YB}
    \begin{abstract}
        We consider the question of whether it is worth building an
        experiment with the sole purpose of bringing the detectable limit on the
        tensor-to-scalar ratio $r$ down to $10^{-3}$. We look at the inflationary
        models which give a prediction in this region and recap
        the current situation with the tensor mode, showing that
        there are only three known models of inflation which give definitive
        predictions in the region $10^{-3}<r<10^{-2}$.
    \end{abstract}

    \maketitle

    \section{Introduction}

        Inflation predicts a unique spectrum of
        gravitational waves with a tensor spectral index
        $n_T\lesssim{}0$
        a detection of which
        would provide the ``smoking'' gun that would affirm inflation \cite{smoke}.
        Currently, model fitting of the CMB data is used to
        obtain the allowed range on the tensor-to-scalar ratio $r$
        , the updated fit by Ref.~\cite{wmap3} places
        the upper limit on $r$ from WMAP as $r<0.90$ with no running in the
        spectral index. 
        The PLANCK satellite promises the ability
        to detect $r\gtrsim0.005$ \cite{planck}, and the ground
        based CLOVER claims that it will bring this limit down to
        $r\gtrsim10^{-2}$ \cite{clover}.


        There are in fact two methods to measure the primordial
        gravitational wave background (pGWB), either indirectly
        for large scales via the B-mode of the CMB-polarisation
        such as ESA's B-mode satellite \cite{bsat} or NASA's inflation
        probe \cite{probe}, or directly for small
        scales using space based laser interferometers such as
        NASA's Big Bang Observer (BBO) or Japan's Deci-Hertz
        Interferometer Gravitational Wave Observatory (DECIGO). The
        contribution of these experiments to deciphering
        inflation's signatures have been considered
        \cite{Smith:2006xf,Smith:2005mm,EC,Boyle:2005se}.
        Refs.\cite{Smith:2005mm,Smith:2006xf} use both model
        dependent and independent methods to evaluate the pGWB
        amplitude predicted by inflation, they conclude that direct
        detection methods would not further constrain models, but
        that if $r\lesssim{}0.01$ would be useful in constraining
        the scale of inflation. Ref.~\cite{Boyle:2005se} concur with
        this and further note the degeneracy in measuring the tensor
        power spectrum at small (BBO) scales between models
        which predict notably different tensor power spectra at
        large (CMB) scales. However they do point out that depending
        on results of the BBO, there is potential to gain deeper
        insight into the physics of reheating. We note that this
        does have (indirect) implications in constraining
        inflationary models, since various models of inflation
        predict various lower limits on the number of $e-$folds of
        inflation when combined with observational limits on the
        spectral index \cite{AL1,AL2}.

        Judging by these papers, any advances in model
        discrimination from the pGWB signal will have to come from
        measuring $r$ at CMB scales, and the onus is on
        CMB-polarisation experiments. However, the future
        experiments focusing on this are the B-mode satellite and
        the inflation probe, with the ultimate lower limit on $r$ being
        $r\sim10^{-3}$, yet we can measure $r\gtrsim10^{-2}$ with
        PLANCK and CLOVER. So the question is whether it is worth
        building new experiments solely to measure
        $10^{-2}<r\lesssim10^{-3}$.

        \par{}This question has been raised for measuring
        $10^{-2}<r<10^{-4}$ in Ref.~\cite{EC}.
        They noted that this level of improvement in the
        detectability of $r$ corresponded to an improvement on the
        limit on the energy scale on the order of $\log(\Delta{}V)\sim4$. They argue
        that there is no reason to expect the energy range to be
        so narrow, and conclude by advising that the next
        generation of CMB probes focus on other signatures of
        inflation, such as the non-gaussian signal.

        \par{}In this paper we reconsider the purpose of such an
        improvement in $r$ detectability by looking at the
        specific predictions for $r$ from models of inflation. We
        revisit the $\log(r) \rm{vs.} n$ plot in Ref.~\cite{AL1}, and add
        the logarithmic bounds on $r$ from a WMAP year 3 
        data fit. We also discuss the assumptions made to obtain
        the $r$ predictions and make amends to the prediction for
        mutated hybrid inflation so that it conforms to
        particle-physics requirements.

        \par{}This paper is laid out as follows, in section \textbf{II} we
        recap the definition of $r$, then in section \textbf{III} we review
        the inflation models giving an $r$ prediction, in section
        \textbf{III A} we re-calculate the upper bound on the $r$
        prediction for the mutated hybrid inflation model, and in section \textbf{III B} we
        briefly introduce brane induced inflation. In
        section \textbf{IV} we discuss the possibility of
        detecting the gravitational wave background (GWB) signature in the region $-3<\log(r)<-2$.

    \section{The Tensor-to-Scalar Ratio}

        Using the convention of Ref.~\cite{leach}, we define the
        gravitational wave contribution of inflation via the ratio
        of the tensor to scalar spectra:

        \be
        r=\frac{\mathcal{P}\sub{grav}}{\mathcal{P}\sub{\mathcal{R}}}=16\epsilon
        \ee
        where $\epsilon=M\sub{pl}^2/2\left(V'/V\right)^2$  is the slow roll
        parameter.
        This can be
        rewritten in terms of the energy scale of inflation, using
        the WMAP year 3 results \cite{wmap3} for the amplitude of the spectrum:

        \[
        r=\left(\frac{V^{1/4}}{3.3\times10^6\rm{GeV}}\right)^4
        \]

        Therefore, $\log(r)$ ranging between $-3<\log(r)<-2$
        corresponds to a tiny range of energy scales
        $\log(\Delta(V))\sim4$. This was noted in Ref.~\cite{EC}, and
        formed the basis for the subsequent discussion and
        conclusion.

        Considering that scales on which the tensor is observed
        leave the horizon after $4$ $e-$folds of inflation \cite{Lyth:1996im}, we can
        use the expression
        $N=M\sub{pl}^{-1}\int_{\phi\sub{end}}^{\phi_{*}}d\phi/\sqrt{2\epsilon}$,
        to relate the tensor fraction to the number of field
        variation over these $4$ $e-$folds:

        \[r=8\left(\frac{\Delta\phi_4}{M\sub{pl}N_4}\right)^2
        \]
        where $\Delta\phi_4$ is the field variation over the $4$
        $e-$folds of inflation, and $N_4\sim4$.
        This expression is known as the \emph{Lyth Bound} \cite{Lyth:1996im}, and if $\epsilon$ increases with time, we get the relation
        between the total field variation $\Delta\phi$ and $r$ \cite{Boubekeur:2005zm}:

        \be\label{field}
            r<0.0032\left(\frac{50}{N}\right)^2\left(\frac{\Delta\phi}{M\sub{pl}}\right)^2
        \ee

    \section{WMAP year 3 bounds}

        We revisit the $\log(r) \rm{vs.} n$ plot of Ref.~\cite{AL1} and
        add the logarithmic limits calculated by a WMAP year 3 
        data fit computed by the authors of Ref.~\cite{wmap3}.
        Note that the prediction for the mutated hybrid
        models is different than the original, the reasoning for
        this appears in the following subsection. We have also
        added the prediction for brane inflation.

        \begin{figure*}
            \includegraphics[angle=270, width=\linewidth]{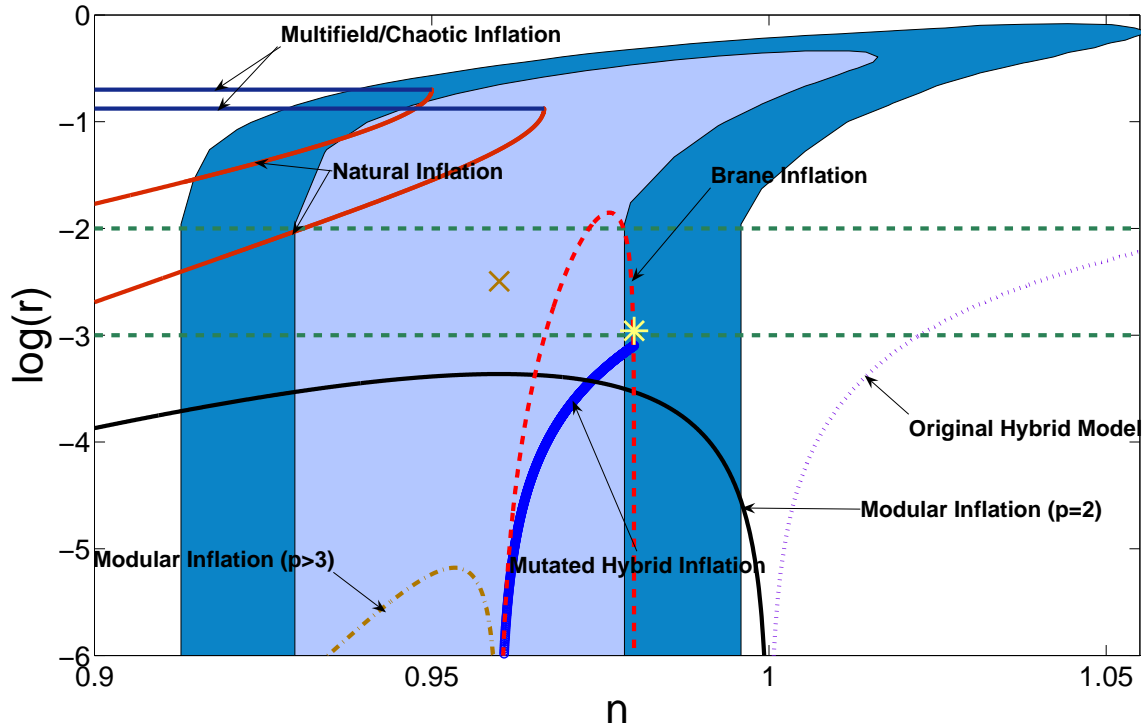}
            \caption{(Colour Online)The updated plot of $\log(r)$ vs. $n$ from Ref.~\cite{AL1}. The shaded regions are the logarithmic fits of the WMAP 
            data set
            at the $1\sigma$ (light blue) and $2\sigma$ (darker blue) confidence levels. The ``$\times$" corresponds to the exponential potential \eq{exp}
            and the ``$\ast$" corresponds to the F/D term inflation model \eq{FD}. We have plotted the predictions for Natural Inflation for $N=40$
            (upper red line) and $60$ (lower red line). The horizontal dashed (green) lines mark the boundaries of the region of interest.
            Note that this region is quite empty (of predictions), thus the motivation for
            dubbing it the ``desert" by David Lyth.}
            \label{logrn}
        \end{figure*}

        \subsection{Mutated Hybrid Inflation}
            In the original analysis we assumed that for the mutated hybrid model, inflation ended at
                $\phi\sub{end}=M\sub{pl}$, which implied that the scale of interest exited the
                horizon at $\phi_{*}>M\sub{pl}$. This
                is stretching effective field theory
                to beyond any reasonable limits, and we recalculate the upper limit on $r$.
                Recalling that the
                potential for such a model is:

                \be\label{mutant}
                    V=V_0\left[1-\left(\frac{\phi}{\mu}\right)^p\right]
                \ee
                we now take inflation to end at
                $\phi\sub{end}\ll{}M\sub{pl}$ 
                . In this case we assume
                that $\phi_{*}\sim{}M\sub{pl}$ and thus that the
                overall field variation is on the order of the
                Planck mass.
                Using the
                equation for the number of $e$-folds $N$ \cite{AL1}:

                \[
                N=\left.\frac{1}{p(p-2)}\left(\frac{\mu}{M\sub{pl}}\right)^2\left(\frac{\mu}{\phi_{*}}\right)^{p-2}\right\vert_{p<0}
                \]
                we set $M\sub{pl}=\phi_{*}=1$ to get:

                \[
                \mu=\left[Np(p-1)\right]^{1/p}
                \]

                Substituting this result into the equation for
                $r$ \cite{AL1}:

                \bea\label{r}
                r&=&16\epsilon\nonumber\\
                &=&8M\sub{pl}^2\left[\frac{p}{\mu}\left(\frac{\phi_{*}}{\mu}\right)^{p-1}\right]^2\nonumber\\
                &=&8\mu^{2p/(p-2)}p^2\left[p(p-2)\right]^{-2\left(\frac{p-1}{p-2}\right)}N^{-2\left(\frac{p-1}{p-2}\right)}\nonumber\\
                &=&8(p-2)^{-2}N^{-2}
                \eea
                in the second step we again set $\phi_*=M\sub{pl}=1$. In Ref.~\cite{AL1}
                it was shown that for this model the spectral index can be related to $p$ via the simple
                relation:

                \be
                    n-1=-\frac{2}{N}\left(\frac{p-1}{p-2}\right)
                \ee
                we find that for $p<0$ (as required by this model) then $n<0.98$,
                and we plot $\log(r)$ from \eq{r} vs. $n$ for $N=50$ in \fig{logrn}.

                In \fig{logrn} we show the prediction for
                $\mu=M\sub{pl}$ and $\Delta\phi=M\sub{pl}$, with
                the latter being a more reasonable assumption for
                this model. Either way, we expect the actual
                prediction to be well bellow these limits, so it
                is not that crucial which one is used.

            \subsection{Brane Inflation}

                \par{}Another model which motivates a power law
                potential equivalent to \eq{mutant} comes from
                M-theory. In this case inflation is induced by the
                separation of branes in extra-dimensions \cite{DT} and the potential is given by:
                \be\label{brane}
                    V=T\left[\alpha+\frac{\gamma}{\phi^{\mathcal{N}-2}}\right]
                \ee
                where $T$ is the tension on the brane(s), $\gamma$ and $\alpha$
                are constants of the model, and $\mathcal{N}$ is the number
                of extra large-dimensions (by large we mean of a
                size $>M\sub{pl}^{-1}$). There are other terms, but according to
                Ref.~\cite{DT} the power law term can dominate, and
                we focus on this for our purposes.

                By comparing the power-law potential of
                \eq{brane} to \eq{mutant}, we find that
                $p=\mathcal{N}-2$, $\mu^p=Np(p-2)\phi_{*}^{p-2}$, and
                has to satisfy \cite{DT}
                $\mu^p<Np(1-p)$.
                We therefore substitute
                the upper limit on $\mu$, which corresponds to $\mu=M\sub{pl}$ into
                \eq{r}, and plot in \fig{logrn}.
        \section{A Closer Look at the Desert}

        The $r$ predictions for the inflation models plotted in
        \fig{logrn} are mostly upper limits. For modular
        inflation ($p=2$ \& $p>3$), mutated hybrid inflation and
        the original hybrid model, the field variations over the
        period of inflation are taken to be $\sim{}M\sub{pl}$.
        For these models to fall within the scope of an effective
        field theory we would require $\Delta\phi<M\sub{pl}$, which means that for $N=50$
        $\log(r)\ll-2.5$. It
        is therefore clear from \fig{logrn} that the tensor
        signatures for these models will probably never be observed.

        \par{}Hope for a detection by PLANCK or CLOVER may be present in the Natural
                Inflation Model, \cite{Freese:1990rb}.
                This is a  sinusoidal potential that motivates
                large field chaotic inflation, and
                which gives a prediction
                in the desert. From \fig{logrn}, note that this prediction lies in the region
                $n<0.94$, thus only accounting for a sliver of the
                region.

        \par{}The most interesting model known which gives a prediction
        in the range of interest is the exponential potential
        , corresponding to the ``$\mathbf{\times}$" in \fig{logrn}. This
        model is given by the potential:

        \be\label{exp}
        V=V_0\left(1-e^{-q\phi/M\sub{pl}}\right)
        \ee
        and can be
        motivated either by a kinetic term passing through zero
        ($q=\sqrt{2}$)
        or non-Einstein gravity ($q=\sqrt{2/3}$)\cite{review}.
        The tensor-to-scalar ratio is then given by:

        \[
        r=\frac{8}{q^2N^2}\left\{
        \begin{array}{cc}
        4/N^2& \rm{Kinetic}\\
        12/N^2&\rm{non-Einstein}
        \end{array}
        \right.
        \]

        which correspond to $\log(r)=-2.80$ and $\log(r)=-2.30$ for $N=50$.

        \par{}Moving on to the ``$\ast$" in \fig{logrn} which sits on $\log(r)=-2.95$ for$N=50$ ,
        which is
        the logarithmic part of F/D-term inflation. The potential for this model is

        \be\label{FD}
        V=V_0\left(1+\frac{g^2}{8\pi^2}\ln{\left(\frac{\phi}{Q}\right)}\right)
        \ee
        where $g\lesssim1$ is the coupling of the field to the water fall and $Q\sim\phi$. In this case
         $r=0.0011(50/N)^2g^2$, and \emph{may}
        be detectable in the upcoming generation of experiments.

        \par{}Brane inflation also gives a prediction in this
        region for $0.96<n<0.98$, and again we have plotted the
        upper limit. The actual expectation is
        much lower.

        \section{Discussion}
        There are three currently known models which give definitive predictions in
        the desert, five if we take into account F/D-term inflation
        and a near upper-limit of brane inflation; the other
        models either give predictions in the $\log(r)\ll-3$ range
        or will have been ruled in or out by experiments measuring
        up to $\log(r)=-2$. We thus concur with Ref.~\cite{EC} that
        a singular mission to detect the gravitational wave background on angular scales of
        $\sim2^o$ \footnote{$2^o$ is the
        angular scale on which the signature of primordial
        gravitational waves can be found} is impractical, as we
        do not expect a detection in this range.

        \par{}The beyond Einstein inflation probe and B-mode satellite, overcome the
        unpromising nature of detecting the tensor-to-scalar ratio
        by including a wider range of multipoles to get a fuller
        spectrum of the B-mode polarisation signature. This means
        that both the scales on which the primordial gravitational wave background
        and those on which the re-ionization signatures are included, hence enabling a
        wider range of research interests to benefit (see for example Ref.~\cite{KM}), while
        still probing the tensor desert.


        \par{}On a final note, this particular signature of inflation is crucial in
        extending the work on discriminating between models of inflation \cite{AL1,AL2}.
        If a detection at $\log(r)>-3$ \emph{is} made, all models
        with $\Delta\phi<M\sub{pl}$ would be eliminated, and even
        if a detection is not made, it would automatically rule
        out models of inflation with
        $\Delta\phi\geqslant{}M\sub{pl}$.
        The latter case would also set a a more definitive upper limit on $\mu$ in brane inflation, as well a slightly stronger
        constraint on the $g$ term in the logarithmic potential.


     \section{Acknowledgments}
        I would like to thank David Lyth for his helpful
        suggestions, and also Joanna Dunkley for kindly supplying me with the
        updated fits for the tensor fraction. I also thank Lancaster University for the award of the studentship from the Dowager Countess
        Eleanor Peel Trust.

\bibliographystyle{unsrt}
    \bibliography{laila2_06}

\end{document}